# Theoretical analysis of reflected ray error from surface slope error and their application to the solar concentrated collector

Weidong Huang

Department of Earth Chemistry and Environmental Science, University of Science and Technology of China

96 Jinzhai road, Hefei, Anhui 230026

Corresponding author: Huang, email: huangwd@ustc.edu.cn

Tel: 86 551 3606631 [O], Fax:86 551 3607386

Abstracts: Surface slope error of concentrator is one of the main factors to influence the performance of the solar concentrated collectors which cause deviation of reflected ray and reduce the intercepted radiation. This paper presents the general equation to calculate the standard deviation of reflected ray error from slope error through geometry optics, applying the equation to calculate the standard deviation of reflected ray error for 5 kinds of solar concentrated reflector, provide typical results. The results indicate that the slope error is transferred to the reflected ray in more than 2 folds when the incidence angle is more than 0. The equation for reflected ray error is generally fit for all reflection surfaces, and can also be applied to control the error in designing an abaxial optical system.

# 光学误差对反射光线影响的理论分析与及其在聚光太阳能系统应用

黄卫东
中国科学技术大学地球化学与环境科学系，
安徽合肥金寨路 96 号科大地学院，230026， email: huangwd@ustc.edu.cn

摘要：聚光反射镜光学误差是影响聚光太阳能系统性能的主要因素之一，光学误差传递到反射光线上，降低了接收器拦截的辐射能量。本文根据几何光学理论，提出了根据反射镜表面误差计算反射光线误差的通用表达式。结果表明，与通常人们认为的简单传递关系不同，在离轴光学系统中，光学误差传递到反射光线上，存在误差放大效应。我们应用该表达式分析了 5 种主要聚光反射系统，给出了反射光线误差的计算表达式和计算结果。在大入射角下，抛物面槽式系统，定日镜，线聚焦塔式系统和固定球面聚光系统的反射光线误差都大于镜面光学误差的两倍，在竖向入射角接近 pi/2 时，线性菲涅耳聚光系统的反射光线误差会增加到光线误差 10 倍以上。。本文提出的反射镜面误差传递公式适用于所有反射面，可以指导离轴光学系统设计中的误差控制。

0、引言

良好的聚光反射镜光学误差较小，能将更多能量聚焦到接收器上(Ulmer, Heinz et al. 2009)。为了减小聚光反射镜光学误差，改进聚光反射镜质量，人们发展了多种方法来测量这种误差(Wendelin and Grossman 1994; Shortis and Johnston 1996; Arqueros, Jimenez et al. 2003; Ulmer, Heinz et al. 2009)。反射镜光学误差的主要方面是光学表面的形状误差，可以分成两种误差，一种是入射位置不同于理论位置，另一种是表面坡度与理论值不同(Winston, Miñano et al. 2005)。从光学角度来看，通常的观点认为位置误差不重要，只需要考虑坡度误差(Rabl 1985)。坡度误差是反射镜表面倾斜角度与理想镜面的差别(Ulmer, Heinz et al. 2009)，一般将这种倾斜角度误差分解成相互垂直的两个方向上的分量，分别用 $dw_x$ and $dw_y$ 来表示，其方向的选取，与镜面加工工艺相关产生的偏差分布有关，可以通过光学测量获得。

根据反射定理，入射光经理想镜面反射后，反射光与法线之间夹角等于入射光与法线之间的夹角，而且入射线，反射线和法线在同一个平面内。在实际镜面上，由于存在倾斜偏差，反射点的法线方向因倾斜方向改变而改变了，反射方向也就改变了。通常镜面各反射点的法线方向的改变是随机的，反射光线的角度分布也是随机，使用分布函数来描述。现有实验数据表明，镜面误差和反射光线的误差分布可用标准椭圆高斯分布描述(Butler and Pettit 1977; Bendt and Rabl 1981)：

$$B_{eff,Gauss}(\theta) = \frac{1}{2\pi\sigma_x\sigma_y}\exp(-\frac{\theta_x^2}{2\sigma_x^2} - \frac{\theta_y^2}{2\sigma_y^2}) \qquad (1)$$

这里 $\theta_x$、$\theta_y$ 是入射角在 x 和 y 方向分量，$\sigma_x$ 和 $\sigma_y$ 是误差分布在两个方向上的标准偏差。通常我们实验测量得到的是特定情况下的反射光线误差，一般认为，光学表面坡度误差引起的反射光线误差是坡度误差的两倍(Winston, Miñano et al. 2005)，从而简单地将表面坡度误差与测量得到的反射光线误差关联起来。然而，实际上它们之间的关系是相当复杂的。例如， Bendt 等给出了抛物槽式聚光太阳能系统两者关系的表达式(Bendt, Rabi et al. 1979)：

$$\sigma_x^2 = 4\sigma_{slopex}^2 + 4(\tan\lambda\sin\frac{\varphi}{2})^2\sigma_{slopey}^2 \qquad (2)$$

这里 $\lambda$ 是入射光线的入射角，$\varphi$ 是槽式系统边缘角，$\sigma_{slopex}$ 和 $\sigma_{slopey}$ 分别是反射面在 x 和 y 方向上的倾斜误差。Bendt 的结果表明，反射镜光学误差传递到反射光线上，不仅与两个方向上的光学误差相关，还与入射角和反射点边缘角相关。Harris 和 Duff 给出了点聚焦和线聚焦聚光太阳能系统的高斯分布参数表达式(Harris and Duff 1981)，其线聚焦槽式系统的表达式虽不同于上式，一样表明了反射镜光学误差传递到反射光线误差，与两个方向上的光学误差以及入射角和反射点边缘角相关。总的来说，如何将光学表面的坡度误差与反射光线误差关联起来，文献上报道较少，目前还没有建立一般情况下的对应关系。

本文根据几何光学原理，从理论上分析了任意一种反射镜的光学误差到反射光线的传递，给出了计算反射镜光学误差引起的反射光线误差的通用表达式,应用该一般表达式分析得到了 5 种常见聚光反射太阳能系统的计算式,计算了典型情况下的结果。

# 1、一般原理

测量光学误差时,通常测量结果描述为镜面法线或镜面倾角在两个相互垂直面上投影的角度,计算他们与理想镜面系统的偏差。对于镜面上任一点 P 来说,假设其法线在两个面上投影与坐标轴 z 轴的夹角分别是 $w_x$,和 $w_y$,则法向量可写成:

$$n = (\tan w_x, \tan w_y, 1) / \sqrt{1+\tan^2 w_x + \tan^2 w_y}; \quad (3)$$

将上式求导,则我们可以得到镜面误差为 $dw_x$,和 $dw_y$ 时的法向量误差,将坐标系的 z 轴与理想镜面法向量重合时,结果可简化为:

$$d\vec{n} = (dw_x, dw_y, 0); \quad (4)$$

根据反射定理我们有:

$$\vec{r} = 2(\vec{i} \cdot \vec{n})\vec{n} - \vec{i} = (-i_x, -i_y, i_z) \quad (5)$$

假设反射线在 xz 面上投影与 z 轴夹角为 $\theta_x$,在 yz 面上投影与 z 轴夹角为 $\theta_y$,则

$$\tan\theta_x = r_x/r_z \quad (6)$$

两边求导,我们得到:

$$d\theta_x = \frac{r_z dr_x - r_x dr_z}{r_x^2 + r_z^2} \quad (7)$$

将(4)和(5)式代入到上式,我们得到

$$d\theta_x = 2dw_x + \frac{2i_x i_y}{i_x^2 + i_z^2} dw_y = 2dw_x + \sin(2\lambda_x)\tan\lambda_y dw_y \quad (8)$$

同理我们得到:

$$d\theta_y = 2dw_y + \frac{2i_x i_y}{i_y^2 + i_z^2} dw_x = 2dw_y + \sin(2\lambda_y)\tan\lambda_x dw_x \quad (9)$$

这里 $\lambda_x$ 和 $\lambda_y$ 是入射光线在 x 方向和 y 方向分量,i 是入射线方向矢量。从推导过程可以看出,(9)和(8)是根据光学误差计算反射光线方向误差的一般表达式,适用于所有反射面。从两式可以看出,反射光线偏差不仅与镜面光学误差相关,还与入射线方向相关。

需要注意的,在应用(9)和(8)式计算时,我们需要将入射光学矢量换算到所在点的法线为 z 轴的坐标系上,x 轴和 y 轴是由镜面误差测量的 x 和 y 方向确定。

如果镜面在两个方向上误差的分布相互独立,标准偏差分别是 $\sigma_{slopex}$, $\sigma_{slopey}$,则反射光线投影在两个方向上的偏差可按下式计算:

$$\sigma_x^2 = 4\sigma_{slopex}^2 + (\frac{2i_x i_y}{i_x^2 + i_z^2}\sigma_{slopey})^2 = 4\sigma_{slopex}^2 + (\sin(2\lambda_x)\tan\lambda_y)^2 \sigma_{slopey}^2 \quad (10)$$

$$\sigma_y^2 = 4\sigma_{slopey}^2 + (\frac{2i_x i_y}{i_y^2 + i_z^2}\sigma_{slopex})^2 = 4\sigma_{slopey}^2 + (\sin(2\lambda_y)\tan\lambda_x)^2 \sigma_{slopex}^2 \quad (11)$$

从式10和11可以看出，在离轴光学系统中，当入射光线与x或y方向不平行时，镜面在两个方向的光学误差都会传输到反射光线的两个方向上。根据（10）和（11）式，对于近轴共轴光学系统来说，$i_x*i_y$很小，式中后面一项可忽略，也就是说，镜面倾斜误差传递到反射光线上加倍了。 但是，对于离轴光学系统来说，通常$i_x*i_y>0$，式中后面一项就不能忽略，说明这时的反射光线在x方向误差， 不仅来自反射镜面x方向倾斜误差，而且y方向误差也部分传递到反射光线上，从而增加了反射光线误差。由于式中后面一项在x或y方向入射角接近pi/2时，可取值很大，说明高入射角下，光学系统误差传递到反射光线上，可增加到非常大的程度。

2、槽式系统

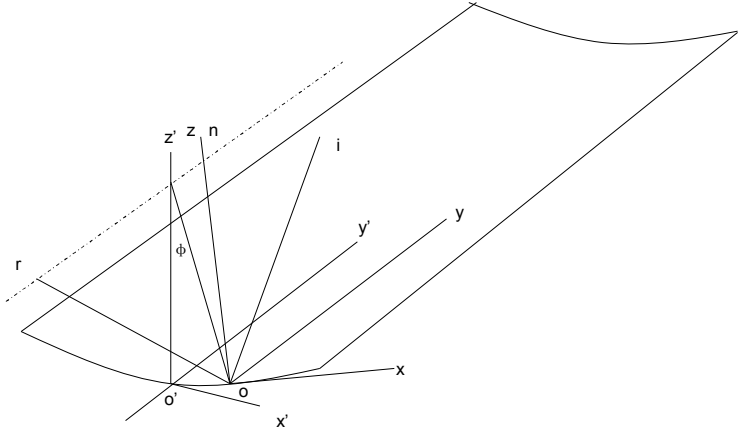

图1 槽式系统反射光线误差计算

槽式系统(Pitz-Paal, Dersch et al. 2007; Grena 2010)是聚光太阳能的主要技术之一。如图1所示，对槽式系统来说，以对称轴为z轴，长度方向为y轴的x'y'z'坐标系内，入射光线矢量为：

i'＝(0，sinλ，cosλ)                                    (12)

对镜面上任一点O来说，假设其边缘角为φ，则将x'y'z'坐标系绕y'轴旋转－φ/2，再平移到O点，就可以得到以O点法线为z轴的Oxyz坐标系上，这时入射光线矢量在Oxyz坐标系下可表示为：

i＝(cosλsin(φ/2),sinλ,cosλcos(φ/2))；                 (13)

对于线聚焦槽式系统来说，我们只需要关心反射光线在x方向上的偏差，我们将上式代入到（10）式得到：

$$\sigma_x^2 = 4\sigma_{slopex}^2 + 4(\tan\lambda\sin\frac{\varphi}{2})^2\sigma_{slopey}^2 \qquad (14)$$

上式与Bendt得到的结果(Bendt, Rabi et al. 1979)是完全一致的，这验证了我们得到的一般表达式。该结果表明，在抛物槽聚光系统中，在入射角大于0时，

两个方向的光学误差都传递到反射光线上了。

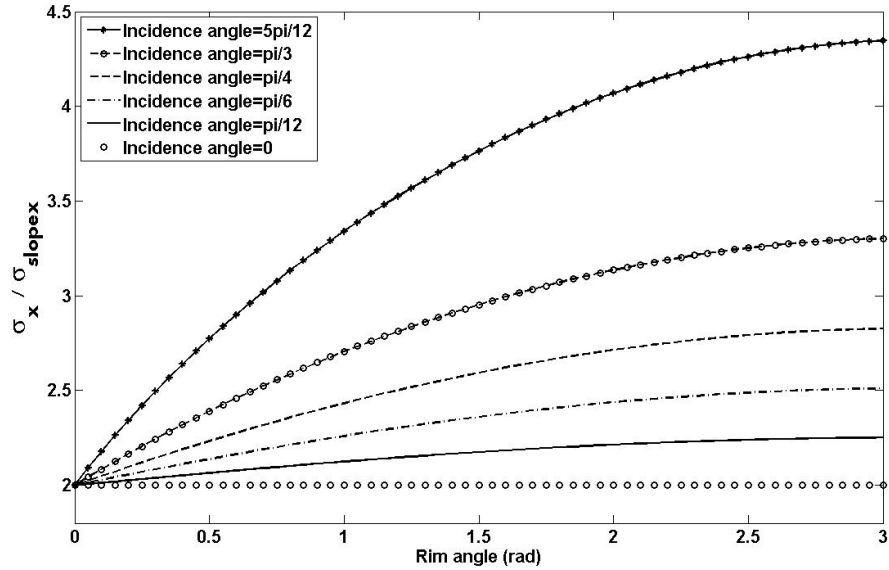

图 2 抛物面槽式系统光学误差到反射光学误差的传递，假设光学误差在 x 和 y 方向相同

图 2 给出了抛物面槽式系统光学误差到反射光学误差的传递的计算结果，计算中假设光学误差在 x 和 y 方向相同。结果表明，同样边缘角下，x 方向反射线误差随入射角增大而增大；一定入射角下，x 方向反射线误差随边缘角增大而增大。这是由于 y 方向光学误差传递到 x 方向，使 x 方向反射线误差增大。从结果可以看出，在入射角为 pi/3 时，反射光学误差可增大 50%以上。

3、定日镜

定日镜(Yao, Wang et al. 2009)通常是用球面制作的，由于焦距常比通光直径大得多，满足 F/D>15，可以近似看成是旋转抛物面镜。首先我们考虑入射线法线所在平面与镜面相交线上的点，对任意一定 P 来说，其法线与球面对称轴的夹角 α，可按下式计算：

$$\sin\alpha = x/r \quad (15)$$

这里 x 是该点到球面对称轴距离，r 是球面半径，或等于旋转抛物面镜焦距的 2 倍。以 P 点法线作为 z 轴，入射线和法线所在平面为 xz 平面的坐标系下，入射光线矢量为：

$$i_P = (\sin(\lambda-\alpha), 0, \cos(\lambda-\alpha)) \quad (16)$$

对定日镜上任意一点 Q 来说，可以将球面对称轴为 z 轴，球面中心 O 为原点的坐标系绕 z 轴旋转 β，使直线 OQ 在 xyo 平面内，这时入射光线在此坐标轴下

$$i = (\sin\lambda\cos\beta, \sin(\lambda)\sin\beta, \cos(\lambda)) \quad (17)$$

再将此坐标系绕其 y 轴旋转 −α，平移到 Q 点，则得到的坐标系是 Q 点为原点，Q 点法线为 z 轴。在此坐标系下，入射光线矢量为：

$$i = (\sin\lambda\cos\beta\cos\alpha - \cos\lambda\sin\alpha, \sin(\lambda)\sin\beta, \sin\lambda\cos\beta\sin\alpha + \cos\lambda\cos\alpha) \quad (18)$$

对定日镜来说，通常通光半径远小于曲面半径，所以 α 很小，这里近似认为 sinα＝0；cosα＝1。所以，我们得到：

$$i = (\sin\lambda\cos\beta, \sin\lambda\sin\beta, \cos\lambda) \quad (19)$$

将上式代入到（10）和（11）式，我们得到：

$$\sigma_x^2 = 4\sigma_{slopex}^2 + (\frac{2(\tan\lambda\cos\beta\cos\alpha - \sin\alpha)\tan\lambda\sin\beta}{1+\tan^2\lambda\cos^2\beta})^2\sigma_{slopey}^2$$

$$\approx 4\sigma_{slopex}^2 + (\frac{2\tan^2\lambda\cos\beta\sin\beta}{1+\tan^2\lambda\cos^2\beta}\delta_{slopey})^2 \quad (20)$$

$$\sigma_y^2 = 4\sigma_{slopey}^2 + (\frac{2(\tan\lambda\cos\beta\cos\alpha - \sin\alpha)\tan\lambda\sin\beta}{\tan^2\lambda\sin^2\beta + (\tan\lambda\cos\beta\sin\alpha + \cos\alpha)^2})^2\sigma_{slopex}^2$$

$$\approx 4\sigma_{slopey}^2 + (\frac{2\tan^2\lambda\cos\beta\sin\beta}{1+\tan^2\lambda\sin^2\beta}\sigma_{slopex})^2 \quad (21)$$

图3和4给出了光学误差传递到定日镜反射光线在x方向和y方向的计算结果。其中方位角是反射点所在位置到定日镜中心线与弧矢面夹角。从结果可以看出，在入射角较大时，由于另外一个方向上的光学误差转移，导致反射光线误差明显增大，例如，入射角为pi/3时，最大反射光线在x或y方向误差会增加20%以上。这会明显降低拦截率和光学效率。

当光学误差分布参数在两个相互垂直方向上不相等时，这时的误差分布是正态椭圆分布。通常定日镜光学误差较小，反射光强分布与高斯分布相差较大，需要计算误差分布函数与太阳光强分布函数的卷积。如果光学误差分布是正态椭圆分布时，其计算量大。上述结果显示，当镜面两个方向上的光学误差相等，而且入射角<pi/4时，描述光学系统误差分布函数的参数在两个方向上也相差很小，可以简化为正态一维高斯分布，计算量会大大减小。此时，（20）和（21）式后一项均比较小，反射光线误差近似等于光学误差。

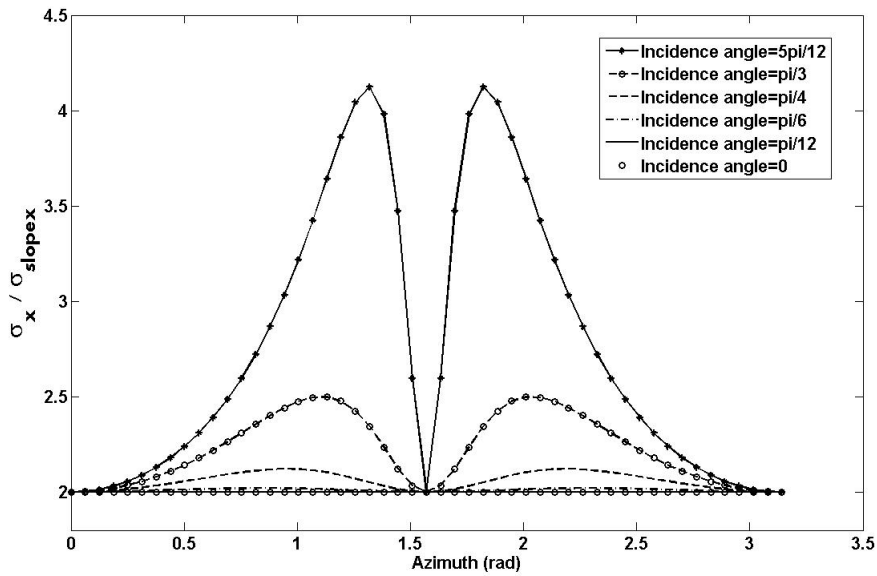

图 3 定日镜反射光线在 x 方向误差与光学误差关系：假设光学误差在 x 和 y 方向相同

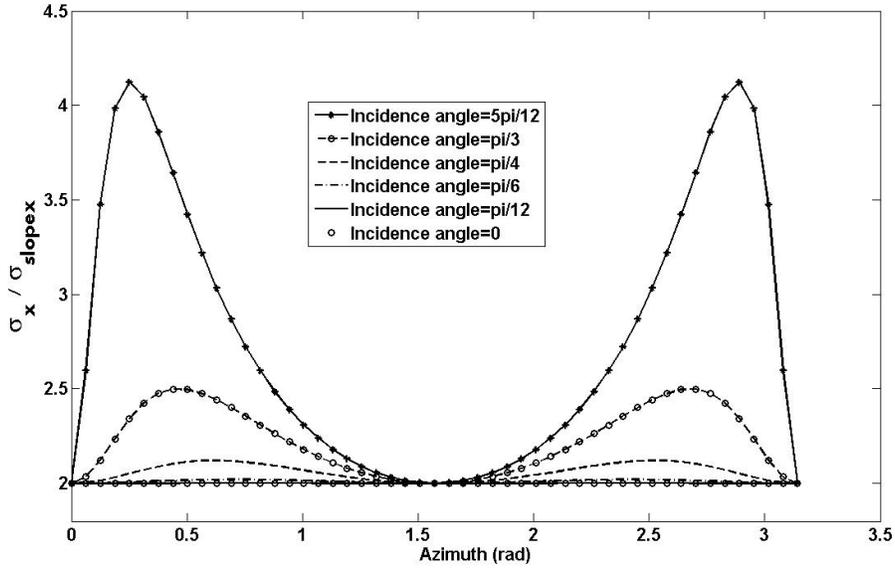

图 4 定日镜反射光线在 y 方向误差与光学误差关系：假设光学误差在 x 和 y 方向相同

4、碟式系统

对旋转抛物面碟式反射镜(Shuai, Xia et al. 2008)上任意一点来说，xz 面是通过该点和主轴的面。对任意一定 P 来说，其法线与对称轴的夹角是 φ/2，所以，以 P 点法线为 z 轴，入射光线矢量为：

$$i = (\sin(\varphi/2), 0, \cos(\varphi/2)),  \quad (22)$$

将其代入到（10）和（11）式，我们得到：

$$\sigma_x^2 = 4\delta_x^2 \quad (23)$$

$$\sigma_x^2 = 4\delta_y^2 \quad (24)$$

这说明，对旋转抛物面碟式系统来说，误差传递是简单的光学误差传递给了反射光线。

通常旋转抛物面镜制造成本高，使用多个小球面镜组合成碟式系统成本较低，受到关注(Johnston, Lovegrove et al. 2003; Riveros-Rosas, Sanchez-Gonzalez et al. 2011)。对每个球面镜来说，可以近似看成是定日镜，入射角 λ 为 φ/2，将其代入到（19）和（20）式，这里 α 也很小，我们将其忽略。所以：

$$\sigma_x^2 = 4\sigma_{slopex}^2 + \left(\frac{2\tan^2(\varphi/2)\sin\beta\cos\beta}{1+\tan^2(\varphi/2)\cos^2\beta}\sigma_{slopey}\right)^2 \quad (25)$$

$$\sigma_x^2 = 4\sigma_{slopey}^2 + \left(\frac{2\tan^2(\varphi/2)\sin\beta\cos\beta}{1+\tan^2(\varphi/2)\sin^2\beta}\sigma_{slopex}\right)^2 \quad (26)$$

该式与定日镜光线误差表达式相同，计算出来的曲线形状，如图 5 和 6，也与定日镜相同。在定日镜中是影响反射光线误差的包括入射角，在球面组合碟式系统中，每个球面的入射角是球面中心所在位置在抛物面上的边缘角的一半。参考对定日镜分析结果，我们可以看出，采取球面组合碟式系统，镜面误差会随球面镜中心在抛物面上的边缘角增加而增加，从而降低了系统性能，说明对球面组合碟式系统来说，使用边缘角较大的系统，性能会因为镜面误差传递到反射光线上的增加而下降，一般不应大于 pi/2。

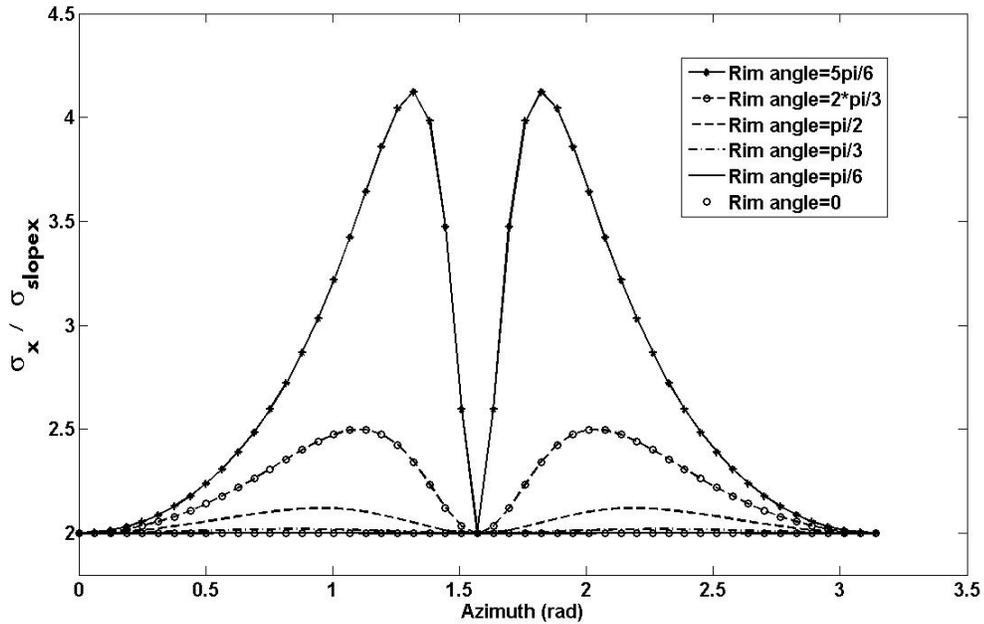

图 5 球面镜组合碟式抛物面反射光线在 x 方向误差与光学误差关系：假设光学误差在 x 和 y 方向相同，边缘角是球面中心在旋转抛物面上的边缘角，方位角是球面上反射点位置到球面中心连线与球面中心与抛物面对称轴面之间夹角。

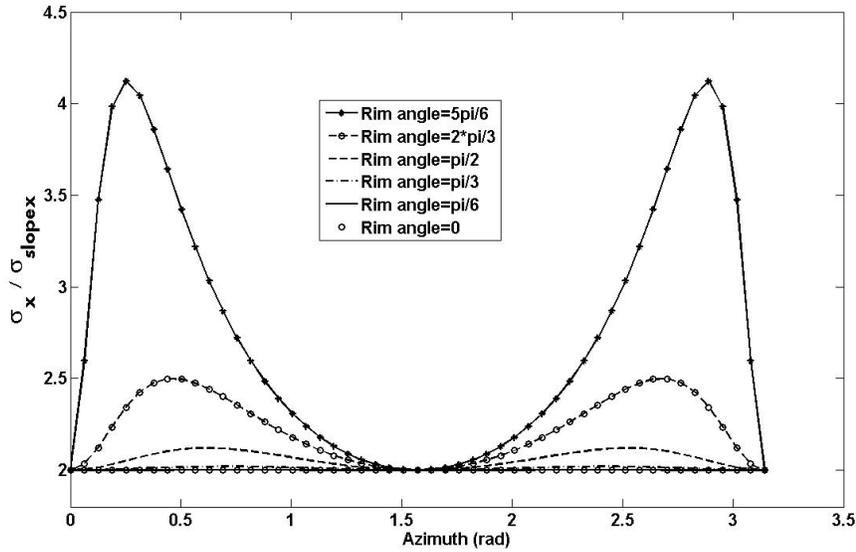

图 6 球面镜组合碟式抛物面反射光线在 y 方向误差与光学误差关系：假设光学误差在 x 和 y 方向相同，边缘角是球面中心在旋转抛物面上的边缘角，方位角是球面上反射点位置到球面中心连线与球面中心与抛物面对称轴面之间夹角

5、线聚焦塔式系统

通常线聚焦塔式系统常使用菲涅耳反射镜(Mills and Morrison 2000)，也可使用圆柱面镜，这里讨论使用正面镀反射膜的圆柱面镜。假设入射光线在系统长度方向上入射角为 λy，在另一个方向上的入射角为 λx，则对镜面上任一点 P 来说，以镜面对称轴为 z 轴的坐标系下，入射光线矢量为：

i＝（tan(λx), tan(λy), 1）/sqrt(tan2(λx)+tan2(λy)+1)    (27)

以 P 点法线为 z 轴的坐标系是上述坐标系绕 y 轴旋转－α，所以，在这个坐标系下，入射光线矢量为：

i = ( tan(λx)cosα - sinα, tan(λy), tan(λx)sinα+cosα ) /sqrt(tan2(λx)+tan2(λy)+1)   (28)

对于线聚焦塔式系统来说，通常焦距很大，而镜面开口宽度较小，可以近似认为：sinα＝0，cosα＝1。所以我们得到：

i = ( tan(λx), tan(λy), 1 ) /sqrt(tan2(λx)+tan2(λy)+1)    (30)

将其代入到（10）式，可得：

$$\sigma_x^2 = 4\sigma_{slopex}^2 + (\sin(2\lambda_x)\tan\lambda_y)^2 \sigma_{slopey}^2 \qquad (31)$$

假设光学误差在 x 和 y 方向相同情况下，根据上式得到的计算结果如图 7。我们可以看到，随着 y 方向入射角增大，反射光线误差增加； 在相同 y 方向入射角下，随着 x 方向入射角的增加，误差会增大，但当 x 方向入射角大于 pi/4 后，又随 x 方向入射角增大而减小，最大误差出现在 x 方向入射角为 pi/4. 通常线聚焦塔式系统安装在南北方向跟踪轴，在高纬度地区，冬天中午太阳的高度角较小，例如，在北纬 40 度地区，冬至日中午太阳高度角仅 26.55º，这时最大误差为：

$$\sigma_x^2 = 4\sigma_x^2 + (\tan[(90-26.55)*pi/180]\sigma_y)^2 = 4\sigma_x^2 + 4.00\sigma_y^2 \qquad (32)$$

这个结果说明，对高纬度地区冬季来说，反射光线误差增大，使性能明显下降。

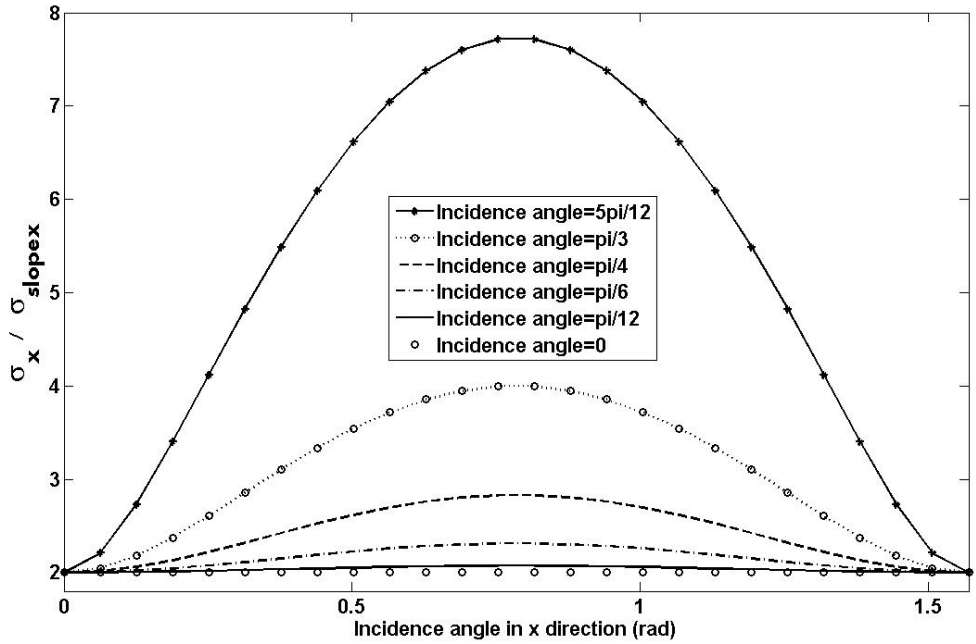

图 7 线聚焦塔式系统光学误差对反射线 x 方向误差的传递，假设光学误差在 x 和 y 方向相同，图中入射角是 y 方向入射角

6、固定球面聚光

球面聚光镜可以固定安装(Sulaiman, Oo et al. 1997)，从而成本较低，受到人们重视。与定日镜情况类似，假设入射角是 $\lambda$，则光学误差产生的反射光方向误差与定日镜的计算式是相同的。其不同的地方在于，由于球面边缘角很大，表达式不能简化，只能使用原式，结果如下：

$$\sigma_x^2 = 4\sigma_{slopex}^2 + (\frac{2(\tan\lambda\cos\beta\cos\alpha - \sin\alpha)\tan\lambda\sin\beta}{1+\tan^2\lambda\cos^2\beta}\sigma_{slopey})^2 \tag{33}$$

$$\sigma_y^2 = 4\sigma_{slopey}^2 + (\frac{2(\tan\lambda\cos\beta\cos\alpha - \sin\alpha)\tan\lambda\sin\beta}{\tan^2\lambda\sin^2\beta + (\tan\lambda\cos\beta\sin\alpha + \cos\alpha)^2}\delta_{slopex})^2 \tag{34}$$

需要注意的是，对于固定球面聚光，当入射角大于 pi/6 时，光线会经过多次镜面反射才能向接收器方向运动，这时需要计算每次反射前的入射光线矢量,按照上述过程多次重复计算每次反射给反射光线带来的方向误差，而相应的反射方向误差就会随反射次数的增加而增加，而经过多次反射的光线，由于误差已经增加到很大了，拦截效率就会下降很多。

7 讨论与结论

本文根据几何光学理论，推导了任意镜面上光学误差传递给反射光线产生的误差计算通用公式，公式表明，传递到反射光线上的误差，不仅与镜面光线误差

相关，而且与入射光线方向相关。对离轴光学系统来说，反射光线误差明显增加，大于系统光学误差。当入射光线角度很大而且与测量时的 x 或 y 方向不平行的时候，光学误差传递到反射光线上，给反射光线带来很大误差。

  我们应用该计算式分析了 5 种主要反射聚光系统情况，包括抛物面槽式系统，塔式系统定日镜，旋转抛物面碟式系统，线聚焦塔式系统和固定球面聚光系统。还包括使用多个球面镜组成的碟式系统，得到了镜面光学误差传递到反射光线上的误差计算表达式和计算结果。 这些结果显示，在大入射角下，抛物面槽式系统，定日镜，线聚焦塔式系统和固定球面聚光系统的反射光线误差都大于镜面光学误差，使用球面组合组成抛物面槽式系统，在边缘角较大时，离系统主轴较远的球面镜反射光线误差也会大大增加。需要注意的是，本文讨论的反射镜是正面镀反射层的反射镜，不是玻璃银镜，没有折射面影响。

  离轴反射光学系统是近年来受到广泛重视而快速发展的技术，它具有较大的自由度，可解决很多领域技术上有较多限制的问题(Geyl 1994)，例如，在同步辐射加速器应用方面的光学系统设计上，为解决光学材料和真空等方面的技术限制，在一些设计中，就存在大入射角光线，从而存在镜面误差传递到反射光线放大问题，例如，非共轴掠入射 KB 与 KBA x 射线显微光学系统(Hu, Zhao et al. 2006)。在这样的光学系统设计中，考虑光学误差传递到反射光线上的放大效应，是非常必要的，将有助于改善系统成像质量。